\documentclass[12pt,a4paper]{article}
\usepackage{cite}
\usepackage[a4paper,top=2cm,bottom=2cm,left=2.5cm,right=2.5cm,marginparwidth=1.75cm]{geometry}
\usepackage{amsmath}
\usepackage{graphicx}
\usepackage[colorlinks=true, allcolors=blue]{hyperref}
\usepackage{hyperref}
\usepackage[title]{appendix}
\usepackage{mathrsfs}
\usepackage{makecell}
\usepackage{amsfonts}
\usepackage{booktabs} % For \toprule, \midrule, \botrule
\usepackage{caption}  % For \caption
\usepackage{threeparttable} % For table footnotes
\usepackage{algorithm}
\usepackage{algorithmicx}
\usepackage{algpseudocode}
\usepackage{listings}
\usepackage{enumitem}
\usepackage{chngcntr}
\usepackage{booktabs}
\usepackage{lipsum}
\usepackage{subcaption}
\usepackage{authblk}
\usepackage[T1]{fontenc}    % Font encoding
\usepackage{csquotes}       % Include csquotes
\usepackage{diagbox}
\usepackage{comment}
\usepackage{url}
\usepackage{graphicx}
\usepackage{lmodern}
\usepackage{multirow}
\usepackage{tabularx}
\usepackage{booktabs}
\usepackage[justification=centering]{caption}
\usepackage[justification=raggedright, singlelinecheck=false]{caption}
\usepackage{helvet}  % Load Helvetica (Arial-like)
\usepackage{lmodern}  % Ensures default LaTeX font is available

\usepackage{setspace}
\usepackage{float}
\usepackage{titlesec}
\titleformat{\section} % Redefine section numbering format
  {\normalfont\Large\bfseries}{\thesection.}{1em}{}

\usepackage{float}   % for customizing caption position
\usepackage{caption}


\date{}  % Remove date

\begin{document}

%----------

\begin{center}
{\LARGE Chemical applicability of $\pi$-Sombor index} \\[5ex]
{\large Monika S} \\
{\small
Department of Physical Science \\
Kristu Jayanti (Deemed to be University) \\
Bangalore, Karnataka, India \\
24csmb33@kristujayanti.com \\
%\url{https://orcid.org/0000-0001-9864-8937}
}  
{\large H. M. Nagesh} \\
{\small
Department of Science and Humanities \\
PES University, Bangalore, Karnataka, India \\
nageshhm@pes.edu \\
%\url{https://orcid.org/0000-0001-9864-8937}
}

\end{center}

\begin{abstract}        
The Sombor index ($SO$) and the modified Sombor index (${}^m SO$) are two closely related vertex-degree-based graph invariants. Both were introduced in the early 2020s and have found diverse applications in chemical, physicochemical, and network-theoretical studies. In 2023, the $\pi$-Sombor index was introduced as the product $SO \cdot {}^m SO$, with an initial focus on its fundamental mathematical properties. However, its potential chemical applicability has remained unexplored. In the present work, we address this gap by employing regression models to analyze the $\pi$-Sombor index in 22 benzenoid hydrocarbons. This study investigates the chemical significance of the $\pi$-Sombor index through curvilinear regression analysis against the physicochemical properties of these hydrocarbons. %\lipsum[1]
\end{abstract}


%\textbf{MSC}: 05C07, 05C09, 05C90, 92E10, 05C92.

\section{Introduction} 
An ordered pair $G = (V,E)$ represents a graph with non-empty vertex set $V(G)$ and edge set $E(G)$. Throughout this article, we only consider a simple, connected, and undirected graph that does not contain a self-loop and multiple edges. The degree of the vertex $u$ is described as $d(u)= |N(u)|$, where $|N(u)|$ is the total number of neighbors of vertex $u$. An arbitrary edge $e \in E(G)$ is represented as $e = uv$ or $vu$. For more information, the readers are referred to \cite{1}. 

Chemical graph theory addresses the details contained within the arrangement of molecules, a factor that significantly influences their physiochemical attributes. Despite a molecule’s topology, as represented by its molecular graph, being a non-quantitative mathematical entity, numerous observable aspects of molecules are typically presented in numerical terms. To establish a connection between molecular topology and these molecular characteristics, the details encoded in the molecular graph need to be quantified. \newpage

The topological index, also known as a molecular descriptor, is crucial in determining this link with no experimental procedures. This index is highly valuable in QSPR analysis, contributing to computer-assisted drug design \cite{2,3}. In short, a topological index is a graph invariant that characterizes the structural properties of a molecular graph. Since Wiener's seminal contribution \cite{4}, researchers have developed numerous types of indices based on different parameters like degree, distance, eccentricity, and spectrum. Since 1970, however, degree-based descriptors have become central to the field, drawing significant interest in both theory and applications.

The Sombor index is a vertex-degree-based graph invariant introduced in 2021 \cite{5}, which eventually gained much popularity. Its mathematical properties and applications in chemistry and other areas have been studied in great detail. It is defined as:
$$
SO = SO(G) = \sum_{uv \in E(G)} \sqrt{d(u)^2 + d(v)^2}
$$

The mathematical properties and the various applications of the Sombor index have been studied in \cite{6,7,8,9,10,11,12}. Shortly after the introduction of the Sombor index, its modified version was proposed \cite{13} as follows:

$$^{m}SO = ^{m}SO(G) = \sum_{uv \in E(G)} \frac{1}{\sqrt{d(u)^2 + d(v)^2}}$$

While the Sombor and modified Sombor indices share many analogous characteristics, they also display a variety of opposing behaviors \cite{14}. Motivated by this close relationship, the study in \cite{14} examined the mathematical behavior of their product and called it as the $\pi$-Sombor index. The mathematical representation of the $\pi$-Sombor index of a graph $G$ is given by
\[
\pi SO=\pi SO(G)=
\left[\sum_{uv\in E(G)}\sqrt{d(u)^2+d(v)^2}\right]
\left[\sum_{uv\in E(G)}
\frac{1}{\sqrt{d(u)^2+d(v)^2}}\right]
\]

\subsection{Motivation and Novelty}
The recently introduced $\pi$-Sombor index ($\pi SO$) is defined as the product of the Sombor index and the modified Sombor index \cite{14}. While its fundamental mathematical properties, lower and upper bounds, and extremal graphs have been established, the foundational literature explicitly stated that its correlating properties with regard to physicochemical parameters of chemical substances remain to be studied \cite{14}. Motivated by this direction for future research, this paper evaluates the chemical applicability of the $\pi$-Sombor index. The novelty of this study lies in utilizing a dataset of 22 benzenoid hydrocarbons to predict their physicochemical properties. Furthermore, rather than relying solely on simple linear regression, the framework is expanded to include quadratic and cubic regression models, thereby mapping the precise linear and non-linear predictive capacity of this vertex-degree-based topological index. The dataset of 22 benzenoid hydrocarbons serves as a structural benchmark to evaluate how the multiplicative nature of the index shifts across varying molecular sizes. 

\section{Chemical applicability of $\pi$-Sombor index}

A growing and diverse array of topological indicators currently exists in the literature. While many of these lack a clear chemical interpretation and are primarily studied from a mathematical perspective, a set of desirable attributes has been compiled to transform this substantial pool of candidates into meaningful descriptors. One such critical attribute is the capability to predict the physicochemical properties of molecules \cite{15}. Evaluating the efficiency of a topological index in modeling physicochemical properties involves correlating its structural characteristics with experimental data across a benchmark set of compounds. 

The chemical applicability of the $\pi$-Sombor index is evaluated in this section. This topological descriptor is utilized to construct linear, quadratic, and cubic regression frameworks to model various physical and chemical properties of benzenoid hydrocarbons. These properties include the boiling point ($BP$), $\pi$-electron energy ($E_{\pi}$), molecular weight ($MW$), polarizability ($PO$), molar volume ($MV$), molar refractivity ($MR$), $XLogP3$, heavy atom count ($HAC$), and complexity ($C$). The specific molecular structures of the 22 benzenoid hydrocarbons are illustrated in Fig.1. The experimental values of the physicochemical properties of the 22 benzenoid hydrocarbons were obtained from \cite{16}, while the corresponding $\pi$-Sombor index values were calculated in the present study and are reported in Table 1.

\begin{figure}[hbt!]
    \centering
    \caption{\centering Structure of 22 lower benzenoid hydrocarbons}
    \label{fig:lower_benzenoid}
    \includegraphics[width=0.55\textwidth]{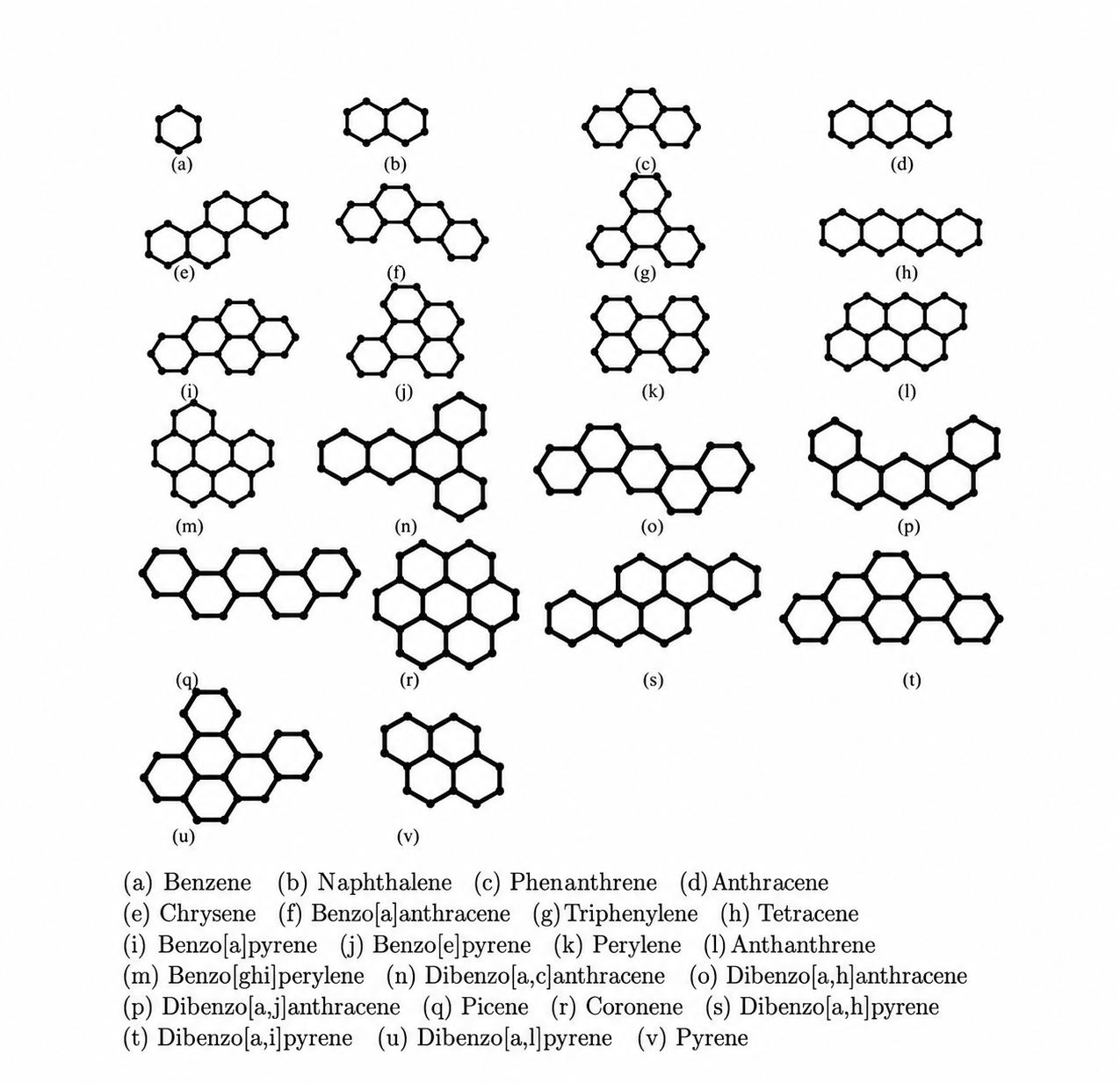} 
\end{figure}

\newpage

\begin{table}[h!]
\centering
\caption{Experimental physicochemical properties and calculated $\pi$-Sombor index values for 22 benzenoid hydrocarbons.}
\label{tab:table 1}
\scriptsize
\setlength{\tabcolsep}{4pt}
\renewcommand{\arraystretch}{2.5}
\begin{tabular}{lrrrrrrrrrrr}
\toprule
Benzenoid hydrocarbons & $\pi SO$ & BP & $E_{\pi}$ & MW & PO & MV & MR & XLogP3 & HAC & C \\
\midrule
Benzene              & 36.0000  & 78.8  & 8.000  & 78.11  & 10.4 & 89.4  & 26.3  & 2.1 & 6  & 15.5 \\
Naphthalene          & 123.5273 & 221.5 & 13.683 & 128.17 & 17.5 & 123.5 & 44.1  & 3.3 & 10 & 80.6 \\
Phenanthrene         & 262.4648 & 337.4 & 19.448 & 178.23 & 24.6 & 157.7 & 61.9  & 4.5 & 14 & 174  \\
Anthracene           & 261.2671 & 337.4 & 19.314 & 178.23 & 24.6 & 157.7 & 61.9  & 4.4 & 14 & 154  \\
Chrysene             & 452.5180 & 448.0 & 25.192 & 228.30 & 31.6 & 191.8 & 79.8  & 5.7 & 18 & 264  \\
Benzo[a]anthracene   & 450.8734 & 436.7 & 25.101 & 228.30 & 31.6 & 191.8 & 79.8  & 5.8 & 18 & 294  \\
Triphenylene         & 454.1531 & 425.0 & 25.275 & 228.30 & 31.6 & 191.8 & 79.8  & 4.9 & 18 & 217  \\
Tetracene            & 449.2190 & 436.7 & 25.188 & 228.30 & 31.6 & 191.8 & 79.8  & 5.9 & 18 & 236  \\
Benzo[a]pyrene       & 590.6940 & 495.0 & 28.222 & 252.30 & 35.8 & 196.1 & 90.3  & 6.0 & 20 & 372  \\
Benzo[e]pyrene       & 592.1548 & 467.5 & 28.336 & 252.30 & 35.8 & 196.1 & 90.3  & 6.4 & 20 & 336  \\
Perylene             & 592.1548 & 467.5 & 28.245 & 252.30 & 35.8 & 196.1 & 90.3  & 5.8 & 20 & 304  \\
Anthanthrene         & 745.1296 & 497.1 & 31.253 & 276.30 & 40.0 & 200.4 & 100.8 & 6.7 & 22 & 411  \\
Benzo[ghi]perylene   & 747.4654 & 501.0 & 31.425 & 276.30 & 40.0 & 200.4 & 100.8 & 6.6 & 22 & 411  \\
Dibenzo[a,c]anthracene & 693.6871 & 518.0 & 30.942 & 278.30 & 38.7 & 225.9 & 97.6 & 6.7 & 22 & 361 \\
Dibenzo[a,h]anthracene & 691.5955 & 524.7 & 30.881 & 278.30 & 38.7 & 225.9 & 97.6 & 6.5 & 22 & 361 \\
Dibenzo[a,j]anthracene & 691.5955 & 524.7 & 30.880 & 278.30 & 38.7 & 225.9 & 97.6 & 6.5 & 22 & 363 \\
Picene               & 693.6871 & 519.0 & 30.943 & 278.30 & 38.7 & 225.9 & 97.6  & 7.0 & 22 & 361  \\
Coronene             & 920.0846 & 525.6 & 34.572 & 300.40 & 44.1 & 204.7 & 111.4 & 7.2 & 24 & 376  \\
Dibenzo[a,h]pyrene   & 861.5507 & 552.3 & 33.928 & 302.40 & 42.9 & 230.2 & 108.1 & 7.0 & 24 & 436  \\
Dibenzo[a,i]pyrene   & 861.5850 & 552.3 & 33.954 & 302.40 & 42.9 & 230.2 & 108.1 & 7.3 & 24 & 436  \\
Dibenzo[a,l]pyrene   & 863.9831 & 552.3 & 34.031 & 302.40 & 42.9 & 230.2 & 108.1 & 7.2 & 24 & 480  \\
Pyrene               & 369.9039 & 404.0 & 22.506 & 202.25 & 28.7 & 162.0 & 72.5  & 4.9 & 16 & 217  \\
\bottomrule
\end{tabular}
\end{table}

The relationship between the $\pi$-Sombor index and the physicochemical properties of benzenoid hydrocarbons was investigated using linear, quadratic, and cubic regression models, the results of which are presented in Tables 2–4. The corresponding scatter plots illustrating these models and their relationships with the physicochemical properties are shown in Fig. 2.

\newpage

\begin{table}[h!]
\centering
\caption{Linear regression models between the $\pi$-Sombor index and physicochemical properties of benzenoid hydrocarbons.}
\label{tab:linear}
\small
\setlength{\tabcolsep}{6pt}
\renewcommand{\arraystretch}{1.9}
\begin{tabular}{lcccc}
\toprule
Linear models & $R^2$ & Adjusted$-R^2$ & $RMSE$ & $SSE$ \\
\midrule
$BP     = 0.4362 \times \pi SO + 200.5307$ & 0.8545 & 0.8472 & 43.2891 & 41226.91 \\
$E_{\pi} = 0.0269 \times \pi SO + 11.6843$  & 0.9461 & 0.9434 & 1.5426  & 52.35    \\
$MW     = 0.2326 \times \pi SO + 110.1304$ & 0.9488 & 0.9462 & 13.0014 & 3718.79  \\
$PO     = 0.0345 \times \pi SO + 14.5382$  & 0.9638 & 0.9620 & 1.6048  & 56.66    \\
$MV     = 0.1340 \times \pi SO + 117.3957$ & 0.8230 & 0.8141 & 14.9508 & 4917.59  \\
$MR     = 0.0868 \times \pi SO + 36.6928$  & 0.9637 & 0.9619 & 4.0547  & 361.70   \\
$XLogP3 = 0.0053 \times \pi SO + 2.8639$   & 0.9227 & 0.9188 & 0.3687  & 2.99     \\
$HAC    = 0.0188 \times \pi SO + 8.5180$   & 0.9522 & 0.9498 & 1.0106  & 22.47    \\
$C      = 0.4743 \times \pi SO + 35.3036$  & 0.9330 & 0.9297 & 30.5585 & 20544.09 \\
\bottomrule
\end{tabular}
\end{table}

\vspace{3mm}

\begin{table}[h!]
\centering
\caption{Quadratic regression models between the $\pi$-Sombor index and physicochemical properties of benzenoid hydrocarbons.}
\label{tab:quadratic_models}
\small
\setlength{\tabcolsep}{6pt}
\renewcommand{\arraystretch}{1.9}
\begin{tabular}{lcccc}
\toprule
Quadratic models & $R^2$ & Adjusted$-R^2$ & $RMSE$ & $SSE$ \\
\midrule
$BP     = -0.000633(\pi SO)^2 + 1.0735(\pi SO) + 79.2096$ & 0.9745 & 0.9718 & 18.1150  & 7219.34  \\
$E_{\pi} = -0.000023(\pi SO)^2 + 0.0505(\pi SO) + 7.1947$  & 0.9941 & 0.9934 & 0.5124   & 5.78     \\
$MW     = -0.000193(\pi SO)^2 + 0.4272(\pi SO) + 73.0894$ & 0.9924 & 0.9916 & 4.9942   & 548.73   \\
$PO     = -0.000024(\pi SO)^2 + 0.0589(\pi SO) + 9.8901$  & 0.9957 & 0.9952 & 0.5534   & 6.74     \\
$MV     = -0.000178(\pi SO)^2 + 0.3136(\pi SO) + 83.2125$ & 0.9202 & 0.9118 & 10.0404  & 2217.82  \\
$MR     = -0.000061(\pi SO)^2 + 0.1487(\pi SO) + 24.9229$ & 0.9958 & 0.9954 & 1.3756   & 41.63    \\
$XLogP3 = -0.000004(\pi SO)^2 + 0.0096(\pi SO) + 2.0374$   & 0.9635 & 0.9596 & 0.2534   & 1.41     \\
$HAC    = -0.000015(\pi SO)^2 + 0.0340(\pi SO) + 5.6147$   & 0.9936 & 0.9930 & 0.3688   & 2.99     \\
$C      = -0.000244(\pi SO)^2 + 0.7199(\pi SO) - 11.4495$  & 0.9495 & 0.9442 & 26.5379  & 15493.71 \\
\bottomrule
\end{tabular}
\end{table}

\newpage
\begin{table}[h!]
\centering
\tiny
\caption{Cubic regression models between the $\pi$-Sombor index and physicochemical properties of benzenoid hydrocarbons.}
\label{tab:cubic_models}
\renewcommand{\arraystretch}{2}
\resizebox{\textwidth}{!}{%
\begin{tabular}{lcccc}
\toprule
Cubic Models & $R^2$ & Adjusted $R^2$ & RMSE & SSE \\
\midrule
$BP     = 0.00000077(\pi SO)^3 - 0.001762(\pi SO)^2 + 1.5263(\pi SO) + 40.1850$ & 0.9844 & 0.9818 & 14.1735 & 4419.52 \\

$E_{\pi} = 0.00000001(\pi SO)^3 - 0.000045(\pi SO)^2 + 0.0593(\pi SO) + 6.4368$ & 0.9951 & 0.9943 & 0.4632 & 4.72 \\

$MW     = 0.00000019(\pi SO)^3 - 0.000470(\pi SO)^2 + 0.5382(\pi SO) + 63.5220$ & 0.9948 & 0.9939 & 4.1585 & 380.45 \\

$PO     = 0.00000003(\pi SO)^3 - 0.000074(\pi SO)^2 + 0.0787(\pi SO) + 8.1833$ & 0.9991 & 0.9990 & 0.2507 & 1.38 \\

$MV     = 0.00000002(\pi SO)^3 - 0.000201(\pi SO)^2 + 0.3226(\pi SO) + 82.4387$ & 0.9202 & 0.9069 & 10.0379 & 2216.72 \\

$MR     = 0.00000009(\pi SO)^3 - 0.000187(\pi SO)^2 + 0.1989(\pi SO) + 20.5971$ & 0.9993 & 0.9992 & 0.5732 & 7.23 \\

$XLogP3 = 0.00000001(\pi SO)^3 - 0.000012(\pi SO)^2 + 0.0128(\pi SO) + 1.7658$ & 0.9670 & 0.9615 & 0.2409 & 1.28 \\

$HAC    = 0.00000002(\pi SO)^3 - 0.000038(\pi SO)^2 + 0.0433(\pi SO) + 4.8173$ & 0.9961 & 0.9955 & 0.2879 & 1.82 \\

$C      = -0.00000041(\pi SO)^3 + 0.000360(\pi SO)^2 + 0.4774(\pi SO) + 9.4466$ & 0.9521 & 0.9441 & 25.8413 & 14690.95 \\

\bottomrule
\end{tabular}
}
\end{table}

\begin{figure}[hbt!]
    \centering    
    \label{fig:lower_benzenoid}
    \includegraphics[width=1.07\textwidth]{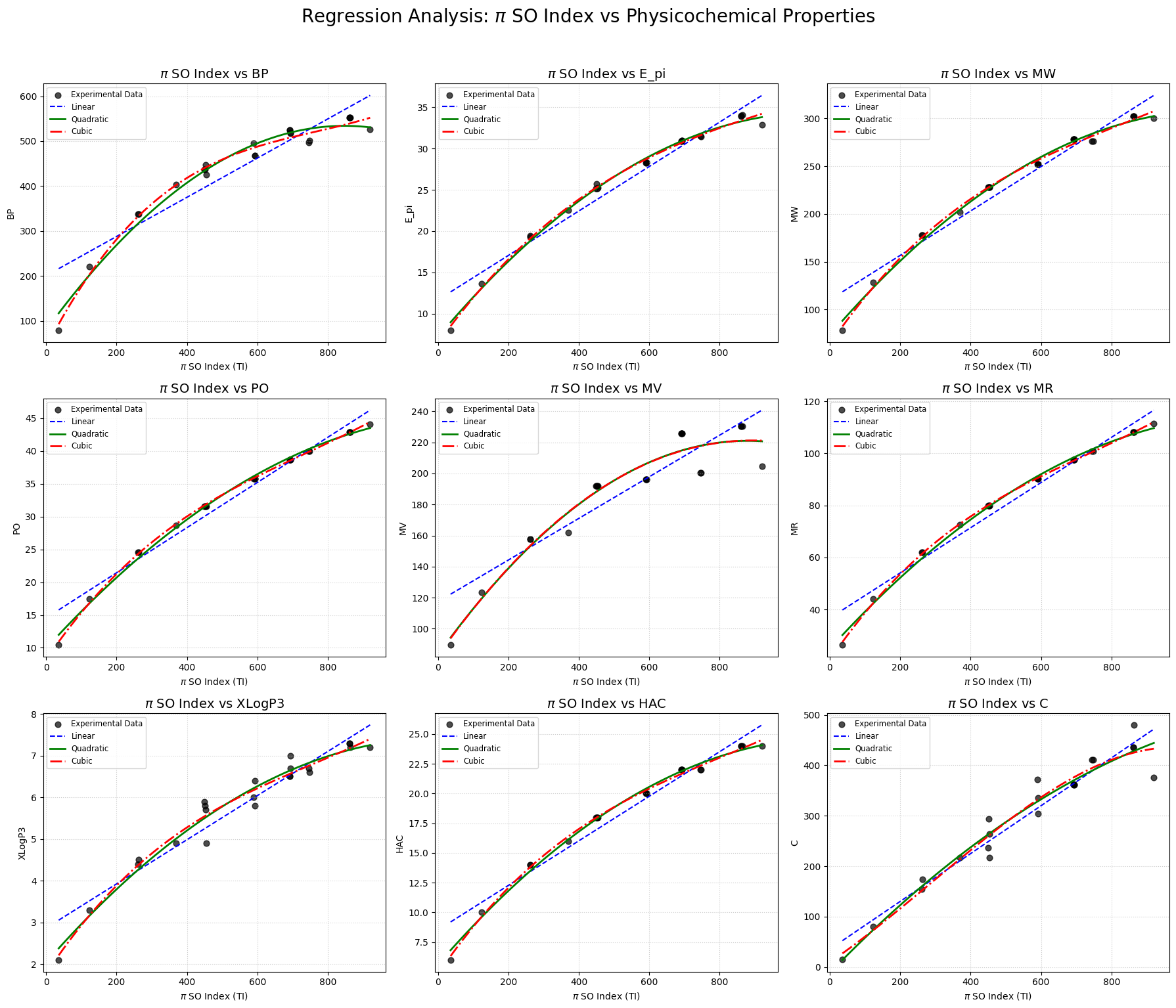} 
    \caption{\centering Scatter plots of $\pi$-Sombor index with physicochemical properties of benzenoid hydrocarbons.}
\end{figure}

\subsection{Interpreting regression model comparison}

To evaluate the predictive power of the $\pi$-Sombor index, we compare the performance of linear, quadratic, and cubic regression models. The ``best” model is chosen based on its ability to maximize the correlation while minimizing the error. We look for the following indicators: High $R^2$: This shows the model is very ``accurate” at following the data points. Low 
$RMSE$ and $SSE$: These represent the ``mistakes” or errors the model makes. The smaller these numbers, the better the model. 

A comprehensive comparison of the statistical parameters across the linear, quadratic, and cubic regression models for the three best predictive properties—namely polarizability ($PO$), molar refractivity ($MR$), and heavy atom count ($HAC$)—is provided in Table 5.

\begin{table}[h!]
\centering
\renewcommand{\arraystretch}{1.3}

\begin{tabular}{llcccc}
\hline
\textbf{Physicochemical property} & 
\textbf{Regression model} & 
\textbf{$R^2$} & 
\textbf{Adjusted-$R^2$} & 
\textbf{RMSE} & 
\textbf{SSE} \\
\hline

\multirow{3}{*}{Polarizability ($PO$)} 
& Linear    & 0.9638 & 0.9620 & 1.6048 & 56.66 \\
& Quadratic & 0.9957 & 0.9952 & 0.5534 & 6.74  \\
& Cubic     & 0.9991 & 0.9990 & 0.2507 & 1.38  \\
\hline

\multirow{3}{*}{Molar Refractivity ($MR$)} 
& Linear    & 0.9637 & 0.9619 & 4.0547 & 361.70 \\
& Quadratic & 0.9958 & 0.9954 & 1.3756 & 41.63  \\
& Cubic     & 0.9993 & 0.9992 & 0.5732 & 7.23   \\
\hline

\multirow{3}{*}{Heavy Atom Count ($HAC$)} 
& Linear    & 0.9522 & 0.9498 & 1.0106 & 22.47 \\
& Quadratic & 0.9936 & 0.9930 & 0.3688 & 2.99  \\
& Cubic     & 0.9961 & 0.9955 & 0.2879 & 1.82  \\
\hline

\end{tabular}

\caption{Comparison of statistical parameters among the curvilinear regression models and the best predictive properties: polarizability ($PO$), molar refractivity ($MR$), and heavy atom count ($HAC$) for the $\pi$-Sombor index.}

\label{tab:comparison}

\end{table}

As shown in Table 5, while the linear model offers a strong initial approximation, the cubic model consistently delivers the highest $R^2$ values and the lowest residual errors, establishing it as the superior framework for predicting the physicochemical behavior of benzenoid hydrocarbons. Among the evaluated parameters, polarizability ($PO$) and molar refractivity ($MR$) emerge as the most highly correlated properties with the $\pi$-Sombor index. Under the cubic regression model, their $R^2$ and adjusted-$R^2$ values near perfection, reaching 0.9991 and 0.9993, respectively. This near-perfect correlation is further substantiated by minimal error metrics, with $RMSE$ values dropping to 0.2507 ($PO$) and 0.5732 ($MR$), and $SSE$ values reducing to 1.38 ($PO$) and 7.23 ($MR$). Heavy atom count ($HAC$) likewise displays excellent predictive performance, achieving an $R^2$ of 0.9961 along with exceptionally low errors ($RMSE = 0.2879$, $SSE = 1.82$) in the cubic formulation. Ultimately, the systematic improvement in these statistical parameters across successive polynomial tiers confirms that higher-order regression models offer a significantly more precise representation of the structural relationship between the $\pi$-Sombor index and these target properties.

Overall, the analysis demonstrates that the cubic regression model provides the most reliable predictive framework for all three properties. Therefore, the $\pi$-Sombor index can be regarded as an effective molecular descriptor for modeling the physicochemical properties of benzenoid hydrocarbons, further supporting its potential applications in QSPR studies.

\subsection{Comparison Analysis}

Based on the statistical parameters obtained from all regression models, the $\pi$-Sombor index is found to be an effective predictor of the physicochemical properties, namely polarizability (PO), molar refractivity (MR), and heavy atom count (HAC), as presented in Table 5.

The obtained results indicate that the $\pi$-Sombor index exhibits improved predictive performance compared with the Elliptic Sombor index reported in \cite{17} for the selected physicochemical properties of benzenoid hydrocarbons. In particular, the cubic regression models based on the $\pi$-Sombor index attain $R^2$ values of 0.9991 and 0.9993 for polarizability (PO) and molar refractivity (MR), respectively, which are higher than the corresponding value of 0.994 reported for the Elliptic Sombor index in \cite{17}. Moreover, the associated RMSE and SSE values are considerably lower, indicating a stronger agreement between the predicted and experimental data. 

The obtained results further demonstrate that the $\pi$-Sombor index effectively captures the structural characteristics of benzenoid hydrocarbons and provides a robust framework for modeling their physicochemical properties. Therefore, the $\pi$-Sombor index can be regarded as a valuable topological descriptor with promising applications in QSPR investigations and chemical graph theory.
\section{Conclusion} 
This study explored the chemical applicability of the recently introduced $\pi$-Sombor index using curvilinear regression analysis on 22 benzenoid hydrocarbons. The results reveal that the $\pi$-Sombor index exhibits significant correlations with several physicochemical properties, demonstrating its potential as a useful molecular descriptor in QSPR studies. These findings highlight the relevance of the $\pi$-Sombor index in chemical graph theory and motivate further investigations on larger and more diverse classes of compounds.\\
\textbf{Some potential limitations:} The present study is subject to certain limitations. First, the analysis is based on a dataset consisting of 22 benzenoid hydrocarbons, which may restrict the generalizability of the obtained results. Second, the investigation is limited to a single class of compounds, and therefore the predictive capability of the $\pi$-Sombor index for structurally diverse molecular families remains to be established. Third, only regression-based relationships with physicochemical properties were considered, without comparison with other established topological descriptors. \newpage Consequently, further studies involving larger datasets, diverse chemical structures, and comparative analyses are necessary to fully assess the applicability and predictive power of the $\pi$-Sombor index in QSPR modeling.\\
\textbf{Future work:} Future research may investigate the applicability of the $\pi$-Sombor index to larger and more structurally diverse classes of chemical compounds. Comparative studies with other established topological descriptors and its incorporation into advanced QSPR/QSAR models may further clarify its predictive capabilities and chemical significance.

\section*{Funding} No funding is available for this study.
\section*{Author contributions}  Monika S: Methodology, verification, original draft writing. Nagesh H. M: Conceptualization, validation.  
\section*{Data and Software Availability}
The data supporting the findings of this study are available within the text and are properly cited via the corresponding references. Software used: Python and MATLAB.
\section*{Declarations}
\textbf{Ethical Approval} Not applicable.\\
\textbf{Conflict of interests} The authors declare that they have no known competing financial interests or personal relationships that could have appeared to influence the work reported in this paper.


\begin{thebibliography}{9}
\bibitem{1} D. B. West, Introduction to graph theory, 2nd edition. (Prentice Hall, Upper Saddle River, 2000).
\bibitem{2} S. C. Basak, A. K. Bhattacharjee, Computational Approaches for the Design of Mosquito Repellent Chemicals, Curr. Med. Chem. \textbf{27(1)} (2020) 32–41. http://dx.doi.org/10.2174/0929867325666181029165413.
\bibitem{3}  S. C. Basak, M. G. Vracko, Parsimony Principle and its Proper use/ Application in Computer-assisted Drug Design and QSAR, Curr. Comput. Aided Drug Des. \textbf{16(1)} (2020) 1–5.  http://dx.doi.org/10.2174/157340991601200106122854.
\bibitem{4} H. Wiener, Structural Determination of Paraffin Boiling Points, J. Am. Chem. Soc. \textbf{69(1)} (1947) 17–20. https://doi.org/10.1021/ja01193a005.
\bibitem{5} I. Gutman, Geometric Approach to Degree–Based Topological Indices: Sombor Indices, MATCH Commun. Math. Comput. Chem. \textbf{86} (2021) 11-16. 
\bibitem{6} H. Liu, I. Gutman, L. You, Y. Huang, Sombor index: review of extremal results and bounds, J. Math. Chem. 60 (2022) 771–798. https://doi.org/10.1007/s10910-022-01333-y.
\bibitem{7} H. Liu, L. You, Y. Huang, Sombor index of c-cyclic chemical graphs, MATCH Commun. Math. Comput. Chem. \textbf{90} (2023) 495–504. https://doi.org/10.46793/match.90-2.495L.
\bibitem{8} K. Naz, S. Ahmad, E. Bashier, On computing techniques for Sombor index of some graphs, Math. Probl. Eng. \textbf{2022} (2022) Article ID 1329653. https://doi.org/10.1155/2022/1329653.
\bibitem{9} M. R. Oboudi, Mean value of the Sombor index of graphs, MATCH Commun. Math. Comput. Chem. \textbf{89} (2023) 733–740. https://doi.org/10.1016/j.amc.2024.128647.
\bibitem{10} M. R. Oboudi, On graphs with integer Sombor index, J. Appl. Math. Comput. \textbf{69} (2023) 941–952.  https://doi.org/10.1007/s12190-022-01778-z.
\bibitem{11} J. Rada, J. M. Rodríguez, J. M. Sigarreta, General properties on Sombor indices, Discrete Appl. Math. \textbf{299} (2021) 87–97. https://doi.org/10.1016/j.dam.2021.04.014.
\bibitem{12} S. Reja, A. Nayeem, On Sombor index and graph energy, MATCH Commun. Math. Comput. Chem. \textbf{89} (2023) 451–465. https://doi.org/10.1016/j.exco.2024.100158.
\bibitem{13} Y. Hunag, H. Liu, On the modified Sombor indices of some aromatic compounds, Journal of South China Normal University (Natural Science Edition). \textbf{53(4)} (2021) 91–99.   https://doi.org/10.6054/j.jscnun.2021063.
\bibitem{14} I. Gutman, I. Red\v{z}epovi\'{c}, B. Furtula, On the product of Sombor and modified Sombor indices, Open J. Discret. Appl. Math. \textbf{6(2)} (2023) 1–6. https://doi.org/10.30538/psrp-odam2023.0083.
\bibitem{15} M. Randi\'c, N. Trinajsti\'c, In search for graph invariants of chemical interest, J. Mol. Struct. \textbf{300} (1993) 551–571.  https://doi.org/10.1016/0022-2860(93)87047-D.
\bibitem{16} J. Barman, S. Das, Chemical applicability of the hyperbolic Sombor index, Chemical Physics Letters. \textbf{878} (2025) 142340. https://doi.org/10.1016/j.cplett.2025.142340.  
\bibitem{17} M. Shanmukha, A. Usha, V. Kulli, K. Shilpa, Chemical applicability and curvilinear regression models of vertex-degree-based topological index: Elliptic Sombor index, Int. J. Quantum Chem. \textbf{124}(9) (2024) e27376. https://doi.org/10.1002/qua.27376.


\end{thebibliography}
\end{document}